\documentclass[runningheads]{llncs}
\usepackage{graphicx}
\usepackage{subfigure}

\usepackage{tikz}
\usepackage{comment}
\usepackage{amsmath,amssymb} 
\usepackage{color}

\begin{document}
\pagestyle{headings}
\mainmatter
\def\ECCVSubNumber{7316}  

\title{Operational vs Convolutional Neural Networks for Image Denoising} 

\titlerunning{ONN vs CNN for Image Denoising}
%
\author{Junaid Malik\inst{1} \and
Serkan Kiranyaz\inst{2} \and
Moncef Gabbouj\inst{1}}
\authorrunning{J. Malik et al.}
%
\institute{Tampere Universities, Tampere, Finland \and
Qatar University, Doha, Qatar
}
\maketitle

\begin{abstract}
Convolutional Neural Networks (CNNs) have recently become a favored technique for image denoising due to its adaptive learning ability, especially with a deep configuration. However, their efficacy is inherently limited owing to their homogenous network formation with the unique use of linear convolution. In this study, we propose a heterogeneous network model which allows greater flexibility for embedding additional non-linearity at the core of the data transformation. To this end, we propose the idea of an operational neuron or Operational Neural Networks (ONN), which enables a flexible non-linear and heterogeneous configuration employing both inter and intra-layer neuronal diversity. Furthermore, we propose a robust operator search strategy inspired by the Hebbian theory, called the Synaptic Plasticity Monitoring (SPM) which can make data-driven choices for non-linearities in any architecture. An extensive set of comparative evaluations of ONNs and CNNs over two severe image denoising problems yield conclusive evidence that ONNs enriched by non-linear operators can achieve a superior denoising performance against CNNs with both equivalent and well-known deep configurations.
\dots
\end{abstract}

\section{Introduction}

Image denoising is a critical low-level computer vision tasks that aims at recovering the original image from its noisy counterpart, which has been corrupted with certain noise. It is an important step in many practical imaging applications, as noise is inherent to most of the digital image acquisition processes. Because the source of noise in real-world images cannot be accurately ascertained, an additive White Gaussian noise (AWGN)-based noise model is generally assumed. The AWGN removal problem has been an active topic for decades. The earliest works mostly viewed it as a local or spectral-domain averaging problem \cite{lindenbaum1994a}. Performance was significantly advanced by the introduction of non-local class of methods, which exploit patch-wise self-similarity in natural images. Especially, the method of BM3D \cite{dabov2007a} held the state-of-the-art status for almost a decade. As with numerous imaging applications, the advent of artificial neural networks, especially Convolutional Neural Networks (CNNs), has shifted the focus towards supervised learning.

CNNs have improved the state-of-the-art in numerous challenging computer vision problems including object recognition \cite{krizhevsky2017a}, detection \cite{he2017a} and segmentation \cite{badrinarayanan2017a}. They are configured as deep artificial neural networks composed of stacked layers of convolutional neurons, each filtering local regions (receptive fields) of the input using the discrete convolution operation with learnable filter coefficients. Such a transformation is especially efficient for large grid-structured data where dense-connectivity is prohibited due to computational requirements. Usually CNN architectures are deep i.e. consisting of several layers, each transforming the previous layer feature maps by convolving them with distinct learnable filterbanks. Moreover, the performance of CNN models is observed to be directly correlated with their depth (number of layers) in the architecture \cite{conneau2017a}. Despite their wide-scale adoption, we identify two key drawbacks that are the primary bottlenecks in CNNs performance, explaining why deep architectures become imperative for practical problems. Firstly, the convolutional neuron model, being an extension of a simple perceptron, is inherently linear, where the lone source of non-linearity is the point-wise non-linear activation. Consequently, a large number of layers with interlaced non-linear activation functions are required to yield a hypothesis space strong enough to navigate complex non-linear spaces. Secondly, while considerable effort has been made to design problem-specific architectures, contemporary CNN architectures are still homogenous i.e. all neurons are identical with the only usage of linear convolution. This can potentially limit the expressiveness of the features extracted by such layers and also prohibits the incorporation of problem-specific neuronal diversity. Accordingly, recent studies have reported the following drawbacks \cite{tian2020a}: (1) it is very difficult to train a deep CNN for denoising tasks, and (2) most of deeper CNNs suffer from performance saturation. Moreover, it is a well-known fact that the existing CNN-based methods cannot work very well on those images with high-level noise \cite{zou2019a}.

In this study, in order to remedy the drawbacks identified in prevalent CNN architectures on image denoising, we first propose a novel heterogeneous network architecture, the Operational Neural Networks (ONN) composed of neurons which allows the flexibility to embed any non-linear operator at the core of the data transformation. This inherently allows diversification of the hypothesis space without the need for increasing the number of trainable parameters. Moreover, we propose a novel approach for the selection of the right non-linear operations based on the task at hand, instead of as a manually tuned hyper parameter. To this end, we introduce a fast, robust and architecture independent method for searching optimal non-linear functions for operational neurons, which hinges on the Synaptic Plasticity paradigm \cite{hughes1958a}; the fundamental principle of learning in bio-neurological circuits. To validate our claims, we train both ONN and an equivalent CNN on challenging denoising problems encapsulating severely corrupted images with speckle (multiplicative) and impulse noise. Owing to the nature of the denoising problem, an ideal network should learn discriminative representations from minimal training data, which are then applicable to any number of images, provided the noise model remains the same. Therefore, we employ harsh training constraints in order to pose a challenging learning problem, which could clearly gauge the learning capability of a model. Finally, we compare the training and generalization performance of ONN with an equivalent CNN as well as a well-known and state-of-the-art deep CNN \cite{zhang2017a}. The novel and significant contributions of this study can be summarized as follows:
\begin{enumerate}
\item We present the concept of a flexible neural network model, the ONN, that enables discriminative non-linear data transformations at a granular level of each receptive field.
\item We propose an architecture-agnostic search strategy based on bio-neurological principles of learning, the Synaptic Plasticity Monitoring (SPM), to identify suitable problem-specific non-linear operators to construct ONNs.
\item We provide evidence that an ONN, configured using optimal non-linear operators, significantly outperforms an equivalent CNN with respect to both training performance and generalization on the denoising problem.
\item This is the first study that deals with denoising such a heavy noise which renders the image content incomprehensible by the naked eye. Despite the only other study that attacked a medium-high level AWGN using a very deep CNN with 52 layers, this study focuses on more challenging noise types such as Speckle and Impulse, using an extremely compact ONN composed of two layers and less than 25 neurons. 
\item A novel Back-Propagation (BP) training method is formulated in a vectorized form that allows direct use of parallel computing paradigms to speed up the training process. 
\end{enumerate}

\section{Related Works}

\subsection{Image Denoising}

One of the pioneering works for the AWGN noise removal problem was that of BM3D \cite{dabov2007a}, which exploit the self-similarity property of natural images by forming stacks of matching non-local patches. A collaborative filtering procedure is then applied where the stacks of patches are transformed to a high-dimensional space and denoised by shrinking the transform coefficients, before inverting them back to the original space. The method achieved state-of-the-art results and held this position for over a decade. Recently, as with many computer vision problems, artificial neural networks ushered in an era of discriminative learning for denoising. In \cite{burger2012a}, it was observed that an MLP can be trained to produce competitive results as compared to BM3D. In \cite{zhang2017a}, a deep CNN architecture was proposed that successfully applied batch normalization and residual learning principles to achieve competitive results for various degrees of AWGN. In order to alleviate the assumption of uniform spatial distribution of noise, authors of \cite{zhang2018a} propose to supplement the inputs with an additional noise map, so that the CNN can learn spatially-invariant encodings for denoising. In \cite{tian2020a}, the residual framework adopted in \cite{zhang2017a} is used and extended with batch renormalization and dilated convolutions to address the problems with small mini-batch and limited receptive fields respectively. The authors of \cite{zou2019a} employ a very deep network, composed of 52 layers, with global and local residual framework to tackle high-level AWGN denoising. Generally, the proposed CNN-based methods employ considerably deep architectures and learn from large-scale datasets; consisting of training examples in the order of $10^5$. Moreover, the noise characteristics of input images are generally mild and preserve the contextual information of the image. Therefore, there is a need to evaluate learning models on more severely corrupted images having different noise characteristics.   Furthermore, as of now, there exists scarce literature that explores the possibility of exploiting non-linearity in the realm of CNNs for image denoising. Nevertheless, few efforts have been made recently to address the major drawback of limited non-linearity in prevalent CNNs.

\subsection{Non-Linear Operations in CNN}
Recently, the authors of \cite{wang2019a} introduced the so-called kervolution operation which proposes to incorporate non-linearity in CNNs by the application of patch-wise kernel trick to transform the data to high-dimensional kernel space, employing the same number of trainable parameters. The study presents a series of widely adopted choices of kernel functions. Evaluation is performed by converting well-known CNN architectures to the proposed kervolution-based models and comparing the performance. Reported results show marginal improvement of the proposed KNN over CNN, with manually chosen kernel functions. In \cite{zoumpourlis2017a}, a Volterra series based second-order expansion of the linear transformation is employed, where the first order term corresponds to the basic linear filter, while the quadratic term introduces the non-linearity, albeit with the additional overhead of n(n+1)/2 trainable parameters. Experiments were performed using the Wide ResNet \cite{he2016a} architecture and modifying it by replacing the first convolutional layer with the proposed second-order expansion of the convolution operation. A slight reduction in the test error over benchmark classification datasets was observed. In \cite{chadha2019a}, the authors focus on the introduction of exponential non-linearity in convolutional neural networks. They primarily explore two ways: introducing non-linearity as a data augmentation strategy in the pre-processing or modifying convolution by adding an additional term with trainable exponential weight parameters. One common observation among the limited methods proposed thus far is the lack of heterogeneity or intra-layer neuronal diversity, which is key towards accurately mimicking the responses of complex visual cortical cells. Secondly, there is no concrete methodology that can provide an end-to-end solution for configuring non-linear networks for a given task. A successful development in this regard has been the introduction of a heterogeneous and non-linear network model, called Generalized Operational Perceptrons (GOPs) \cite{kiranyaz2017a,kiranyaz2017b}. GOPs entail the natural diversity that appears in biological neurons and neural networks. Specifically, the diverse set of neurochemical operations in biological neurons (the non-linear synaptic connections plus the integration process occurring in the soma of a biological neuron model) have been modelled by an “operator set” composed of a nodal operator (for synaptic connection), a pool operator (for the integration in the soma), and an activation function (for activation in the axon). As a result, the GOP neuron model becomes a superset of the conventional MLP and was shown to  achieve superior performance in several benchmark deep learning problems, such as nonlinear function approximation, the “Two-Spirals”, the “N-bit Parity” for N$>$10, and the “White Noise Regression”, problems \cite{tran-a,tran2019a,tran2019b,tran2019c}.

\section{Operational Neural Networks}
The primary building block of the proposed Operational Neural Network framework is the operational neuron model which extends the principles of GOPs to convolutional realm. While retaining the favorable characteristics of sparse-connectivity and weight-sharing in CNN, ONN provides the flexibility to incorporate non-linear transformation within local receptive fields without the overhead of additional trainable parameters. In this section, we provide a brief overview of a practical implementation of convolutional operation in CNN before providing detailed forward and backward propagation formulation for ONN.

\subsection{Preliminaries}
\label{sec:prelim}
The primary building block of a convolutional neuron is the 2D discrete convolution operation. The convolution of a 2D image $y \in \mathbb{R}^{M\times N}$ with a filter $w\in\mathbb{R}^{m\times n}$ is given as:
\begin{align}
  x(i,j)= \sum_{u=0}^{m-1} \sum_{v=0}^{n-1} \left( w(u,v)y(i-u,j-v) \right)
\end{align}

Suppose now an alternate formulation of the above operation. For the sake of brevity, unit stride and dilation is assumed and the input is padded with zeros before the convolution operation in order to preserve the spatial dimensions. Firstly, y is reshuffled such that values inside each $m\times n$ sliding block of $y$ are vectorized and concatenated as rows to form a matrix $Y\in\mathbb{R}^{\widehat{M} \times \widehat{N}}$ where $\widehat{M}=MN$ and $\widehat{N}=mn$. We can denote each element of Y as:
\begin{align} \label{eq:2}
    Y(i,j)=\ y(\lnot(i,j))
\end{align}

where $\lnot$ is a matrix that stores the index locations of $y$ in $Y$. This operation is commonly referred to as “im2col” and is critical in conventional GEMM-based convolution implementations. Secondly, we construct a matrix $W\in \mathbb{R}^{\widehat{M}\times\widehat{N}}$ whose rows are repeated copies of $\vec{w}=\ vec(W)\in\mathbb{R}^{mn}$, where $vec(\bullet)$ is the vectorization operator. Each element of W is given by the following equation:
\begin{align}
    W(i,j)= \vec{w}(i)    
\end{align}

The convolution operation can then be represented as
\begin{align} \label{eq:4}
x=vec_{M\times N}^{-1} \left(\sum_{j} (Y \otimes W) \right)
\end{align}

where $\otimes$ represents the Hadamard product, $\sum_i$ is the summation across the dimension $i$. In (\ref{eq:4}), $vec_{M\times N}^{-1}$ is the inverse vectorization operation that reshapes the vector back to an $M\times N$ tensor. The formulation given in (\ref{eq:4}) can now be generically reformulated as follows:
\begin{align} \label{eq:5}
x=vec_{M\times N}^{-1} \left(\phi\left(\psi(X,W\right)\right)
\end{align}

where $\psi(\bullet): \mathbb{R}^{M\times N}\rightarrow \mathbb{R}^{M\times N}$  and $ \phi(\bullet):\mathbb{R}^{\widehat{M} \times \widehat{N}} \rightarrow \mathbb{R}^{\widehat{M}} $ are termed as nodal and pool functions respectively. The convolution operation of (\ref{eq:4}) is now a special form of (\ref{eq:5}) with nodal function $\psi(\alpha,\beta) = \alpha \ast \beta$ and pooling function $\phi(\bullet) = \sum_{i}(\bullet)$ To complete the forward-propagation, a non-linear activation operation f follows the convolution. So, the final output of a convolutional neuron can be expressed as follows:
\begin{align} \label{eq:6}
y=vec_{M\times N}^{-1} f(x)
\end{align}
where $f(\bullet)$ represents the point-wise activation operation.

\subsection{Forward Propagation}
Here, we show the forward propagation-based on the fore-mentioned formulation, through a neuron in a hidden layer l. We assume for clarity that this neuron is connected to a single neuron in the previous layer, whose output is $y_{l-1}\in\mathbb{R}^{M\times N}$.

As illustrated in  Figure \ref{fig:1}a, $w_l\in\mathbb{R}^{m\times n}$ is the weight connecting the two neurons, $Y_{l-1}$ is the reshuffled form of $y_{l-1}$ based on the size of $w_l$ and $W_l\in\mathbb{R}^{\widehat{M} \times \widehat{N}}$  is the broadcasted form of $w_l$, as introduced in section \ref{sec:prelim}. The forward propagation then proceeds as in (\ref{eq:5}) and (\ref{eq:6}) to obtain the current neuron’s output $y_l$.
\begin{align} \label{eq:7}
y_l=vec_{M\times N}^{-1} f\left(\left(\phi(\psi(Y_{l-1},W_l)\right)\right)
\end{align}

\subsection{Back Propagation}
For backward propagation, the flow of gradients through an operational neuron is illustrated in Figure \ref{fig:1}b. Given the derivative of loss with respect to the neuron’s output, $\frac{\partial E}{\partial y_l}$ , we need to find sensitivities with respect to this neuron’s input $y_{l-1}$  and weight $w_l$. Using vectorized forms for notational convenience, the former can be calculated as follows:

\begin{align} \label{eq:14}
\frac{\partial E}{\partial \vec y_{l-1}} = \frac{\partial E}{\partial \vec Z_l} \frac{\partial \vec Z_l}{\partial \vec y_{l-1}}    
\end{align}

where 

\begin{align} \label{eq:16}
\frac{\partial \vec Z_l}{\partial \vec y_{l-1}}(i,m) = 
\begin{cases}
\frac{\partial \psi(\vec Y_{l-1}(\vec \lnot(i),\vec W_{l}(\vec \lnot(i)))}{\partial \vec Y_{l-1}(m)}, & \text{if}\ m = \vec \lnot(i) \\
0, & otherwise
\end{cases}
\end{align}

In (\ref{eq:16}), $\vec \lnot$ is the vectorized form of the matrix $\lnot$ introduced in (\ref{eq:2}). Similarly, we can calculate $\frac{\partial E}{\partial \vec w_l}$ by computing the following product:
\begin{figure}[t]
\hfill
\subfigure[]{\includegraphics[width=6cm]{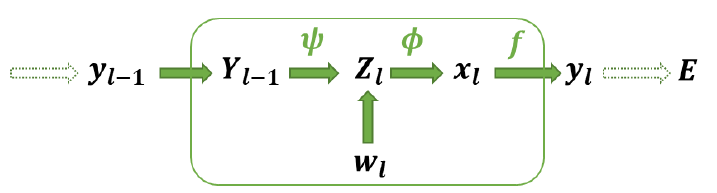}}
\hfill
\subfigure[]{\includegraphics[width=6cm]{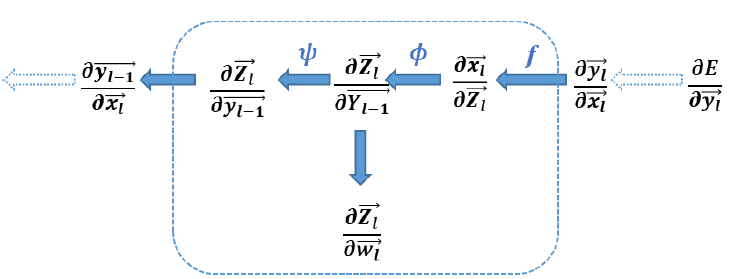}}
\hfill
\caption{Forward (a) and Back-propagation (b) through an operational neuron.}
\label{fig:1}
\end{figure}

\begin{align} \label{eq:17}
    \frac{\partial E}{\partial \vec w_l} = \frac{\partial E}{\partial \vec Z_l}\frac{\partial \vec Z_l}{\partial \vec w_l}
\end{align}
where the Jacobian matrix takes the following form:
\begin{align} \label{eq:18}
    \frac{\partial \vec Z_l}{\partial \vec w_l}(i,m) = 
    \begin{cases}
    \frac{\partial \psi(\vec Y_{l-1}(i),\vec W_l(i))}{\partial \vec w_l(m)} & \text{if}\ m=i \\
    0 & otherwise
    \end{cases}
\end{align}
which can be re-formulated as,
\begin{align} \label{eq:19}
    \frac{\partial \vec Z_l}{\partial \vec w_l}(i,m) = 
    \begin{cases}
    \frac{\partial \psi(\vec y_{l-1}(\vec \lnot(i)),\vec W_l(i))}{\partial \vec w_l(m)} & \text{if}\ m=i\mod|w| \\
    0 & otherwise
    \end{cases}
\end{align}
Detailed derivations of the fore-mentioned equations are presented in the supplementary section.

\subsubsection{Equivalence to CNN}
If the nodal and aggregation function, $\psi$ and $\phi$ are set to multiplication and summation respectively, the formulation of (\ref{eq:7}) collapses to the linear model of a convolutional neuron. The intermediate derivative $\frac{\partial \vec x_l}{\partial \vec Z_l}$ becomes a matrix with all 1’s and $\frac{\partial \vec Z_l}{\partial \vec w_l}$ is a matrix whose $i^{th}$ column is filled with $\vec w_l(i)$. Other more complicated non-linear choices for nodal and pool functions yield more complex formulations for Jacobian matrices.

\subsection{Synaptic Plastic Monitoring (SPM)}
Based on the formulation of Section 3.2, an operational neuron enjoys any choice of first-order differentiable functions $\psi$, $\phi$ and $f$ operators, thus allowing more discriminative non-linear transformations as compared to the fixed model of a convolutional neuron. Furthermore, we observe that application-specific modifications to CNN architectures still assume the fixed linear model of a convolutional neuron and rely on data-driven search for architectural hyperparameters. As a remedy, we propose a fast and robust architecture-agnostic technique to identify optimal choices for $\psi$, $\phi$ and $f$ given any learning problem.

\subsubsection{Hebbian Learning Postulate}
Synaptic plasticity refers to the ability of neurons to shape their inter-association as a response to sensory stimuli in order to learn behavior and drive the process of learning and memory in bio-neurological circuits \cite{hughes1958a}. One of its foundational theories is the Hebbian learning theory or Hebb’s postulate \cite{attneave1950a} which is often summarized as: \textit{neurons that fire together, wire together} \cite{loewel1992a}. Specifically, the strengthening of the synaptic connections between two neurons signals their active involvement in the learning process and strong inter-association. With some interpretational leeway, we can exploit the afore-mentioned rules to rank the efficacy of different operator sets in an operational neural network by linking the suitability of an operator set with the variation in synaptic connection strength that occurs in a pre-synaptic neuron with that operator set.

\subsubsection{Synaptic Plasticity Monitoring}
We refer to two operational neurons in an operational neural network: i) the pre-synaptic neuron $a$ in layer $l-1$ which has been assigned a distinct operator set $\theta_{(i,j,k)}=[\psi_i,\phi_j,f_k]$ and ii) the post-synaptic neuron $b$ in layer $l$. The synapse between these two neurons is represented by the trainable parameter $w_l^{(a,b)}$, which is updated at each learning step based on the gradient descent approach. Therefore, at a time-step t, the weight takes the value $w_l^{(a,b)}(t)$. Given a learning process of $\gamma$ iterations over the training data, the variation in the connection strength can be termed as the health factor, quantified by the change in power (variance) of $w_l^{(a,b)}$. Using this, the synaptic efficacy of a neuron in layer $l-1$  with operator set $\theta_{(i,j,k)}=[\psi_i,\phi_j,f_k]$  is calculated as follows:
\begin{align} \label{eq:20}
\rho^{l-1}(\theta_{(i,j,k)}) = \frac{\left|\sigma^2\left(w_l^{(a,b)} (t)\right) - \sigma^2\left(w_l^{(a,b)} (t-\gamma)\right)\right|}{\left|\sigma^2\left(w_l^{(a,b)} (t-\gamma)\right)\right|}
\end{align}
For a generalized case where neuron $a$ is connected to more than one neuron in the next layer, the powers are averaged before calculating the synaptic efficacy factor.

\subsubsection{Configuration of the “elite” ONN}
Given an architecture $A$ consisting of $L$ layers and a learning problem $P$, the first step is to assign an operator set, chosen randomly from a pre-defined operator set library $\vartheta$, to each individual neuron in $A$. After training the network for $M$ iterations, the health factors of each of the assigned operators for all the layers are calculated as in (\ref{eq:20}), based on which, a distinct ranking of the operator sets’ suitability is obtained for each layer. As the operator assignment and the weight initializations are stochastic, it is a necessary in practice to repeatedly perform the fore-mentioned steps and recalculating $\rho(r)$ at each run $r$, in order to reduce the probability of an operator set suffering from a bad initialization or biased assignment. After a sufficient number of runs $R$ are completed, the final synaptic efficacy is calculated as follows:
\begin{align} \label{eq:21}
\widetilde{\rho^l}\left(\theta_{(i,j,k)}\right) = \sum_{r=1}^{R}\rho^l\left(\theta_{(i,j,k)}\right)(r)\ for \ l=1,..,L-1
\end{align}
We empirically observe, that after a preset number of runs, K, the random assignment can be confined by taking into account the distribution of $\widetilde{\rho^l}\left(\theta_{(i,j,k)}\right)$. Specifically, instead of pure random assignment, we employ the following probability distribution for selection of operator:
\begin{align} \label{eq:22}
P_l(\theta) = \frac{\widetilde{\rho^l}\left(\theta_{(i,j,k)}\right)}{\sum_{\theta_{(i,j,k)}} \widetilde{\rho^l}\left(\theta_{(i,j,k)}\right)}
\end{align}

Finally, after the completion of R runs, we have a layer-wise ranking of optimal operator sets for the given architecture A. To construct the elite ONN, we chose K top ranked operators for each layer and assign operators based on their normalized health factors. Here, K, is the heterogeneity factor, which controls the neuronal diversity within each layer of the constructed ONN. 

\section{Experiments}
\subsection{Datasets}
\begin{figure} 
\centering
\includegraphics[height=5cm]{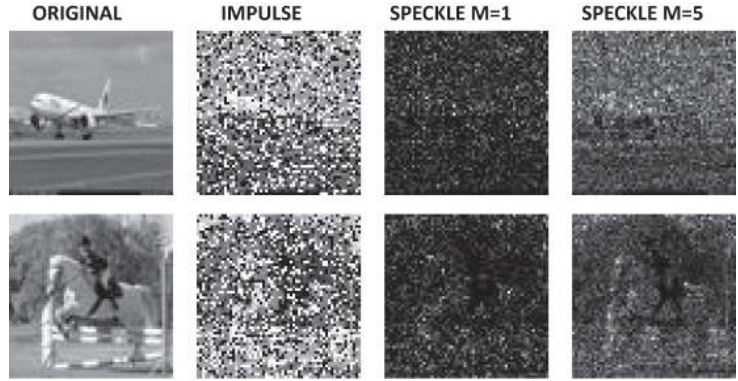}
\caption{Example images from the datasets used in this study.}
\label{fig:3}
\end{figure}

We employ two different types of noise models used to generate the noisy data of 1000 images from the PASCAL \cite{everingham2010a} dataset. All images are converted to grayscale and resized to 60x60 resolution. For impulse noise denoising, we corrupt the images with fixed-value impulse noise with a probability of 0.4. Specifically, in each image, approximately 40\% of the pixel values are randomly replaced with darkest and brightest pixel values possible within the data range. For speckle noise, we employ the model used in \cite{bioucas-dias2010a} for multiplicative noise where the noise probability is given by the Gamma distribution:
\begin{align} \label{eq:23}
p(n) = \frac{M}{\Gamma(M)}  e^{-nM}n^{M-1}    
\end{align}

where the value of $M$ drives the level of corruption. For our experiments, two sets of noisy images corresponding to acute noise levels M=1 and M=5 are used. It is worth noting that such a severe degradation almost entirely wipes out the semantic information of the images and it is difficult to visually derive any meaningful information, as shown in Figure \ref{fig:3}. Hence, this poses an extremely challenging learning problem. For all the problems, we minimize the mean-squared loss and use the Peak Signal-to-Noise Ratio (PSNR) as the evaluation metric:
\begin{align} \label{eq:24}
PSNR(x_{orig},x_{noisy}) = \frac{MAX_{x}^{2}}{\sum_{N}(x_{orig}-x_{noisy})^2}
\end{align}
 
where $MAX_x$ is the maximum possible peak of $x$ in the given data range.

\subsection{Experimental Setup}
\subsubsection{Network Architecture}

Evaluation is performed by comparing 3 models; ONN, a similar size CNN and a deep CNN as proposed in \cite{zhang2017a}. We employ a shallow architecture which is composed of only 2 hidden layers with 12 neurons each. This design choice follows naturally the premise of this study. As a comparison is made between the learning power of ONN and CNN, a shallow network would suffice in gauging the discriminative ability. Nevertheless, it must be noted that ONN and SPM are, by design, generalized and can be applied to any class of neural network architectures. 

\subsubsection{SPM Parameters}
For SPM, we set the hyperparameters $\gamma = 80$, $R=4$ and $K=3$ for all problems. The operator sets are shown in Figure \ref{fig:7}, where mul stands for multiplication, while other operations are well-known non-linear transformations.

For impulse denoising, the optimal operator diversity percentages of $(\psi,\phi)$ pairs identified by SPM was 32\% sinh-sum, 24\% sine-sum, 20\% convolutional, 12\% cubic-median and 8\% cubic-sum, with a common tanh activation function. For speckle noise, the corresponding percentages for $(\psi,f)$  pairs were 24\% log-tanh, 20\% mul-lincut, 16\% cubic-lincut and 12\% each for sine-lincut, cubic-tanh and sine-tanh, with a common choice of summation for the pool function. Hence, the configured ONNs manifested rich heterogeneity, with variety of problem-specific non-linearities incorporated.

\begin{figure}[t]
\hfill
\subfigure[]{\includegraphics[width=4cm]{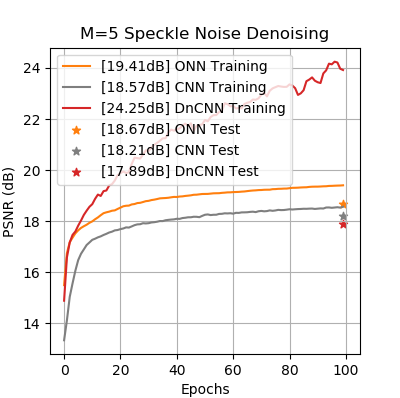}}
\hfill
\subfigure[]{\includegraphics[width=4cm]{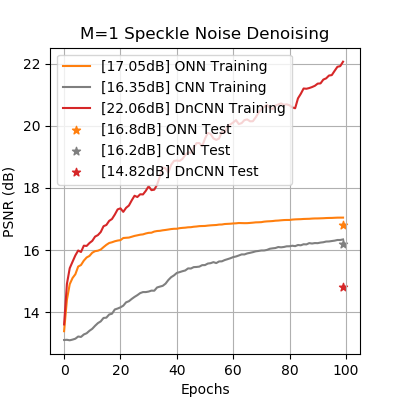}}
\hfill
\subfigure[]{\includegraphics[width=4cm]{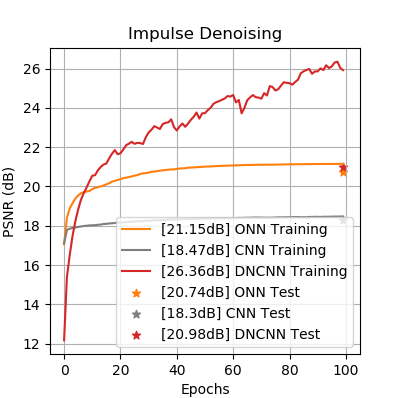}}
\caption{Training and test performance versus the number of epochs for ONN, CNN and DnCNN in the three denoising problems: Speckle Noise with (a) M=5, (b) M=1 and (c) Impulse noise.}
\label{fig:5}
\end{figure}

\subsubsection{Training parameters}
We pose challenging learning constraints in order to assess the learning potential of ONNs versus CNNs. For each problem, a 10-fold cross-validation is applied on the available 1000 images; where the model is trained on 100 images and tested on the remaining 900. Furthermore, we apply early stopping by limiting the number of epochs to 100. For all models, while training, 3 randomly initialized runs are made for each fold and the one which produced the maximum training PSNR is chosen as the representative and is evaluated on the test set.

\subsubsection{Optimization}
The choice of optimization technique is one of the key factors towards efficient training of any architecture and is therefore a widely studied topic. Stochastic Gradient Descent \cite{cauchy1847a} remains the most prevalent technique to train deep neural networks. It involves iteratively updating the trainable parameters; slightly nudging them in the negative direction of the gradients at each step. The earliest vanilla version involves scaling the gradient by a small learning rate before updating the parameter value. While SGD-based optimization generally works well in all cases, adaptive methods have been shown to accelerate convergence by including a measure of the past-gradients in the current step formulation \cite{duchi2010a}. One of the widely adopted method in this regard is ADAM \cite{kingma2015a} which adaptively updates the parameters based on the exponentially weighted averages of first and second moments of their gradients. ADAM has become a de-facto choice for CNN, and is, therefore, used in all our experiments. For ONN, the method was slightly modified based on our empirical observation that using sample variance to estimate the second moment produces more stable results and better convergence.

\section{Results and Discussion}
\begin{figure}[h] 
\centering
\includegraphics[height=5cm]{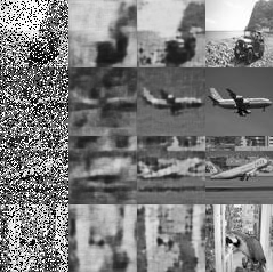}
\caption{Randomly selected noisy (input) images, the corresponding outputs of the CNNs, ONNs and the original (target) image from the test set (from left to right).}
\label{fig:6}
\end{figure} 
\subsection{Training Dynamics}
Figure \ref{fig:5} shows the training plots and test set performance for all three problems considered in this study averaged across all 10 folds. It is apparent that across all applications, ONNs obtain a better convergence compared to the CNNs. We observe a percentage improvement of 14.5\%, 4.28\% and 4.52\% in the training accuracy for the three denoising applications: speckle noise M=1, M=5 and impulsive noise, respectively, as compared to CNNs. In addition to converging to a better local optimum, ONNs also require a smaller number of training steps for achieving the same accuracy level. We observe in our experiments that on average, an ONN requires approximately 80\% less number of training steps to achieve the same PSNR level as a CNN. This trend can be clearly witnessed in Figure \ref{fig:5}. Furthermore, we observe that the deep CNN model achieves the best training accuracy across all three problems. However, it must be noted that the number of trainable parameters in the deep CNN model is 64 times ($>$ 0.5 million) more than of the ONNs ($<$ 9000) used in all experiments. This provides two key insights. First, ONN achieves around 67 times higher performance per trainable parameter as compared to the deep CNN. Secondly, the performance of the CNN with equivalent configuration is increased only 1.36 times when significantly deeper and more complex configuration is used; signaling performance saturation. This shows clear evidence that owing to considerably higher inter- and intra-layer heterogeneity, ONNs are expected to scale significantly better than CNNs.

\subsection{Data-driven Non-Linearity}
\begin{figure}[t]
\hfill
\subfigure[]{\includegraphics[width=6cm]{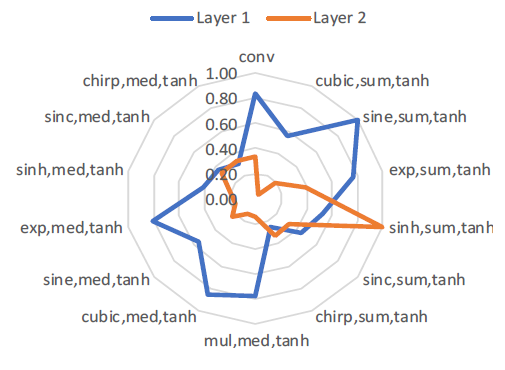}}
\hfill
\subfigure[]{\includegraphics[width=6cm]{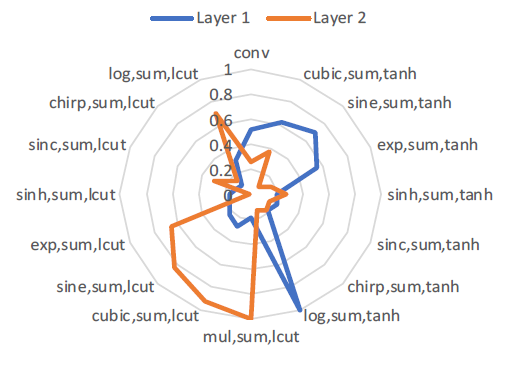}}
\caption{Radar charts showing the normalized health factors for non-linear operators calculated using SPM for Impulse noise (a) and Speckle noise (b).}
\label{fig:7}
\end{figure}

Figure \ref{fig:7} shows the radar chart illustrating the ranking of non-linear operators obtained through SPM for speckle and impulse noise reduction, respectively. It is well known that the median filter is useful for filtering images corrupted with salt and pepper noise \cite{yin1996a}. We see from Figure \ref{fig:7} that SPM for this problem clearly favors the operator sets with median pool function for layer 1. Similarly, for speckle noise, it is common practice to apply a logarithmic transformation in order to obtain an additive noise model \cite{bioucas-dias2010a}. It is apparent from Figure \ref{fig:7} that for speckle noise reduction, SPM favors the log operation in the $1^{st}$ hidden layer. The aforementioned observations validate the notion that SPM identifies suitable non-linear operators in a data-driven fashion.

\subsection{Generalization performance}
Generalization is crucial for any learning-based framework, as it directly correlates with the richness of the learnt feature representations and whether the network is overfitting on the training data. For the denoising application in particular, assuming the same noise characteristics, an ideal model must be able to generalize over any number of images which have been corrupted with the same noise. Table \ref{table:2} details the percentage increase in test set performance (in terms of PSNR) achieved by ONN over competing models. 

\begin{table}[] 
\begin{center}
\caption{Percentage improvement achieved by ONNs over the test set.}\label{table:2}
\begin{tabular}{lccc}
\hline
Network        & Speckle Noise: M=1 & Speckle Noise: M=5 & Impulse Noise \\ \hline
Equivalent CNN   & 2.52\%       & 3.7\%        & 13.33\%       \\
Deep CNN \cite{zhang2017a} & 4.35\%       & 13.36\%      & -1.14\%       \\ \hline
\end{tabular}
\end{center}
\end{table}

It is evidently clear that ONN is highly resistant to overfitting and achieves a significant improvement in PSNR over the competing models in all the addressed denoising problems. This validates the proposed hypothesis that ONNs are much less prone to overfitting and learn rich representations from scarce training data. Figure \ref{fig:6} provides vivid examples of this generalization difference by showing results of the best performing CNN and ONN models on sample images from the test set.

\section{Conclusions}
In this study, we tackle image denoising by proposing Operational Neural Networks (ONN), an important extension of the convolutional neural model. We show that non-linear transformations at the level of individual receptive fields guided by learnable kernels considerably improve the learning ability of the model, as compared to the fixed linear model of CNN. We also propose SPM, a data-driven robust search-strategy that hinges on the principles of Hebbian theory of learning in biological neurons. Results show that SPM is efficient in identifying the right operators given any learning problem. An extensive set of denoising results show that ONNs have a superior generalization ability, even surpassing the performance of the state-of-the-art deep CNN model. An important observation worth mentioning here is that although the best performances of ONNs and CNNs are evaluated by hyper-parameter search with ADAM optimization, ONNs’ performance level reported in this study is still bounded by the limited number of operators used in the sample library. In other words, ONNs performance can further be improved with a richer operation set library, which may encapsulate customized nonlinear operators for denoising certain noise types. Finally, the architecture-agnostic formulation of ONN and SPM naturally lends itself to application on deeper architectures. This will help in gaining further insights into how the biologically inspired non-linear models of artificial neurons can help alleviate the burden of training massive architectures. This will be the primary focus of our future work.

%
%
\bibliographystyle{splncs04}
\bibliography{egbib}
\end{document}